\documentstyle[12pt]{article}
\newcommand{\rf}[1]{(\ref{#1})}
\newcommand{\beq}{\begin{equation}}
\newcommand{\eeq}{\end{equation}}

\renewcommand{\a}{\alpha}

\newcommand{\bea}{\begin{eqnarray}}
\newcommand{\eea}{\end{eqnarray}}

\renewcommand{\ni}{\noindent}

\newcommand{\oh}{\frac{1}{2}}

\newcommand{\vep}{\varepsilon}

\newcommand{\bR}{{\bf R}}

\begin{document}
\topmargin 0pt
\oddsidemargin 5mm
\headheight 0pt
\topskip 0mm

\addtolength{\baselineskip}{0.20\baselineskip}

\pagestyle{empty}
\hfill RH-08-92

\hfill YITP/U-92-31

\hfill September 1992

\begin{center}

\vspace{10pt}
{\Large \bf
On subdivision invariant actions for random
surfaces\footnote{Supported in part by a NATO Science
collaboration grant.}}

\vspace{2 truecm}

{\em Bergfinnur Durhuus}

\medskip
Uji Research Center, Yukawa Institute for Theoretical
Physics\\
Kyoto University, Uji 611\\
Japan
\medskip

and

\medskip

Matematisk Institut, University of Copenhagen \\
Universitetsparken 5, 2100 Copenhagen \O \\
Denmark

\vspace{1.3 truecm}

{\em Thordur Jonsson}

\medskip

Raunvisindastofnun Haskolans, University of Iceland \\
Dunhaga 3, 107 Reykjavik \\
Iceland

\vspace{1.5 truecm}

\end{center}

{\bf Abstract.}  We consider a subdivision invariant
action for dynamically triangulated random surfaces that
was recently proposed \cite{savvidy} and show that it is
unphysical: The grand canonical partition function is
infinite for all values of the coupling constants.  We
conjecture that adding the area action to the action of
\cite{savvidy} leads to a well-behaved theory.

\vfill

\newpage
\pagestyle{plain}

The most straightforward way to discretize the functional
integral for bosonic strings with the Nambu-Goto action
is to sum over all triangulations with the weight $\exp
(-\mu S_A)$ where $S_A$ is the sum of the areas of
individual triangles.  This discretization was studied in
\cite{adf} and the partition function was found to be
divergent due to spikes that are not suppressed by the
weight factor.  Since then the most widely studied random
surface theory is the one with Gaussian action on random
triangulations \cite{adf,d,mig} which is a natural
discretization of the Polyakov functional integral.

The area action has the nice feature that it is
subdivision invariant, i.e. if we have a triangulated
surface imbedded in $\bR ^d$ and refine the triangulation
by adding new vertices and links that lie in the original
surface then the action does not change.  This implies
that two surfaces which are "close to each other" in
imbedding space have almost the same action.  In other
words, the area action is a contionuous
function\footnote{This notion of continuity can of course
be made precise but for our present purposes this is not
necessary.} on the space of imbedded triangulated
surfaces.
The Gaussian action is not continuous in this sense, but
one should keep in mind that it is by no means clear that
such continuity is a correct requirement of a
well-behaved physical theory.

The Gaussian action leads to a string tension which does
not scale to zero at the critical point \cite{ad}, a
pathology which is believed to be caused by the dominance
of branched polymer-like surfaces \cite{dfj,adfo,adf2}.
In \cite{adfj1,adfj2} it was argued that extrinsic
curvature terms should be added to the Gaussian action in
order to prevent the dominance of branched polymers.
Recent numerical work \cite{num1,num2,num3,num4} indicates
that this modification results in a scaling string
tension, but a conclusive proof is still lacking.

In a recent paper \cite{savvidy} a new subdivision
invariant action for random surfaces was proposed.
The purpose of this letter is to point out that this
action suffers from a similar problem as the area action, i.e.
the partition function is divergent. The action of
\cite{savvidy} depends on the angle between
neighbouring triangles, like the extrinsic curvature action
does, but in a subdivision invariant fashion which we now
describe.  Let $T$ be a closed triangulation with one vertex
marked.  A piecewise linear surface $S$ triangulated by $T$ is
given by a mapping $i\mapsto x_i$ from the
vertices of $T$ into $\bR ^d$.  We assume that the marked vertex
$i_0$ is mapped to 0 in order to remove translational
invariance. If $(i,j)$ is a pair of nearest neighbour vertices
in $T$ we denote by $\a _{ij}$ the angle between the planes of
the two triangles in imbedding space which share the link
$(x_i,x_j)$ (i.e. $\a _{ij}=0$ if the triangles lie in the same
plane with the same orientation).
The action for $S$ proposed in \cite{savvidy} is defined by
\beq
A_T=\oh\sum_{(i,j)}|x_i-x_j|\,|\a _{ij}|\label{1}
\eeq
where the sum is over all nearest neighbour pairs of
vertices in $T$.  Since the action associated with a
link vanishes if the adjoining triangles are
parallel and is linear in the length of links, $A_T$ is
readily seen to be subdivision invariant in the sense
described above.

The partition function associated with $T$ is
defined, for $\lambda >0$, by
\beq
Z_T(\lambda )=\int e^{-\lambda A_T}
\prod _{i\in T\setminus \{ i_0\} }dx_i.\label{2}
\eeq
The grand canonical partition function is defined for $\mu >0$
and $\lambda >0$ by
\beq
Z(\mu ,\lambda )=\sum_Te^{-\mu |T|}Z_T(\lambda ),\label{3}
\eeq
where the sum is over all closed triangulations with
a fixed topolgy and $|T|$ is the number of vertices in
$T$.

We now prove that the sum defining $Z(\mu ,\lambda )$ is
divergent for all values of $\mu$ and $\lambda$ and suggest a
modification of the action \rf{1} whose associated grand
canonical partition function we expect to be well defined.  This
divergence is caused by surfaces which are almost planar with a
large distance between vertices. For simplicity we restrict
ourselves to triangulations with spherical topology.  Consider a
square triangulation with $n$ vertices on each edge and regular
inside with all interior vertices of order 6, see Fig. 1.
Next take  an identical square triangulation rotated through
$90^{{\rm o}}$ and glue the two together along their boundaries.

\setlength{\unitlength}{1
truecm}

\begin{picture}(15,6)(-4.8,0)
\multiput(0,0)(0,1){6}{\line(1,0){5}}
\multiput(0,0)(1,0){6}{\line(0,1){5}}
\multiput(0,0)(0,1){6}{\multiput(0,0)(1,0){6}{\circle*{0.1}}}
\multiput(0,0)(0,1){5}{\multiput(0,0)(1,0){5}{\line(1,1){1}}}

\end{picture}
\medskip

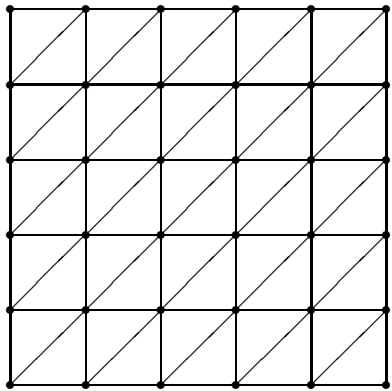
\centerline {\small {\bf Fig. 1:} A regular square triangulation with
6 vertices on each edge.}

\bigskip
\ni
Then we obtain a
closed triangulation of $S^2$ with $2n^2-4n+4$ vertices,
all of which are of order 6, except 4 (the corners) which
are of order 3.  Denote this triangulation by $T_n$ and
let the marked vertex be one of the corners.

Consider the
surface $S_n$ triangulated by $T_n$ defined such that the marked
vertex $i_0$ is mapped to 0 and the other vertices fall on a
planar hexagonal grid, all of whose links have length $a$, in
such a way that the "upper half" of the surface overlaps the
"lower half". Let $B_{n,i}$ be an ellipsoid in $\bR ^d$ centered
on the vertex $x_i$ in $S_n$ with axes of length $a/3$ in the
plane of the surface and axes of length 1 in the $d-2$ coordinate
directions perpendicular to this plane.  We assume that $a\gg 1$
and let $B_n\subset \bR ^{d(|T_n|-1)}$ denote the Cartesian
product over $i$ of all the $B_{n,i}$'s except $B_{n,i_0}$.

We now obtain a lower bound on $Z_{T_n}(\lambda )$ by
restricting the integration domain in \rf{2}, with $T=T_n$, to
$B_n$.  This means that we only take into account perturbations
of the surface $S_n$ for which $x_i\in B_{n,i}$ for $i\in
T\setminus {i_0}$.  For such surfaces we evidently have
$\a_{ij}=O(a^{-1})$, i.e. there is a constant $c_1$ such that
$\a_{ij} \leq c_1/a$, for all links $(i,j)$ except the "boundary
links" where the surface bends back upon itself.  It follows
that
\beq
A_{T_n}\leq 6\lambda (n-2)^2(a+2a/3)c_1a^{-1}+\oh\lambda
(4n-4)(a+2a/3)\pi.\label{4}
\eeq
The first term on the right hand side of \rf{4} is a
bound on the contribution of interior links to the
action while the second term is a bound on the boundary
contribution.  The volume $|B_n|$ of $B_n$ fulfills
\beq
|B_n|\geq (a^2c_2)^{|T_n|-1}\label{5}
\eeq
where $c_2$ is a constant which only depends on $d$.  Hence,
\bea
Z_{T_n}(\lambda )&\geq &(a^2c_2)^{|T_n|-1} \exp \left(-6\lambda
(n-2)^2(a+2a/3)c_1/a
-\lambda (2n-2)(a+2a/3)\pi\right) \nonumber\\
  & \geq & c_3\exp \left((4\ln a -c_4
-c_5\lambda )n^2 -(8\ln a -2c_4+c_5\lambda a)n\right),\label{6}
\eea
where $c_3$, $c_4$ and $c_5$ are positive
constants independent of $n$ and $a$.
{}From \rf{3} and \rf{6} it follows that
\bea
Z(\mu ,\lambda )&\geq &\sum _{n=2}^{\infty}e^{-\mu
|T_n|}\,Z_{T_n}(\lambda )\nonumber \\
  &\geq &c_3'\sum _{n=2}^{\infty}
e^{ (-2\mu +4\ln a -c_4
-c_5\lambda )n^2 -(-4\mu +8\ln a -2c_4+c_5\lambda a
)n},\label{7}
\eea
for a positive constant $c_3'$.  Choosing $a$ sufficiently
large the right hand side of \rf{7} is clearly divergent.
We have therefore shown
that the contribution of a particular class of "smooth" slowly
undulating surfaces gives a divergent contribution to the
partition function.  This could be expected from the
outset since the action $A_T$ does not at all counteract
the entropy of vertices in the plane of the surface.

It is easy to convince oneself that the above
construction can be generalized to surfaces with an
arbitrary boundary and genus.  It also generalizes to
actions of the form
\beq
A_T=\sum _{(i,j)}|x_i-x_j|\theta (\a _{ij})
\eeq
provided
\beq
0\leq\theta (\a )\leq {\rm constant}\,|\a |^{\vep }
\eeq
with $\vep >(d-2)/d$.

It is natural to ask whether it is the unhappy fate of
any subdivision invariant action to lead to a divergent
partition function.
In order to address this question we recall that it was shown in
\cite{adf} that the partition functions associated with a
certain class of triangulations are divergent if the model is
defined by the area action.  More generally, it was proven that
for any triangulation the expectation value of a sufficiently
high power of the mean extent of the surface is divergent.
These divergences are caused by a high probability for the
formation of spikes which extend far out but have small area.
For example, in $d=2$, the integration over a vertex of order 3
gives rise to a nonintegrable singularity in the distances
between the three neighbouring vertices.  On the other hand,
for regular triangulations like $T_n$, the partition function
was not shown to be divergent.  On the contrary, numerical
simulations indicate that it is finite \cite{marinari}.

Thus the following picture emerges: The action \rf{1} clearly
suppresses spiky surfaces.  It is easy to see that the
integration over a single vertex in \rf{2} does not give rise
to a singularity in the distances between the remaining
vertices.  But, as demonstrated above, the action \rf{1} is not
coercive enough to compete successfully with the entropy of
almost flat surfaces, no matter how large we choose the
coupling $\lambda$.  In contrast the area action is well
behaved for regular surfaces but fails to suppress spikes.

On the basis of these observations it does not seem
unreasonable to expect proper thermodynamical behaviour of a
model whose action is a linear combination of the area action
and the action \rf{1}, i.e.
\beq
A'_T=\kappa\sum _{\Delta\in T}|\Delta |+
\lambda\sum _{(i,j)}|x_i-x_j|\,|\a
_{ij}|,\label{11}
\eeq
where the first summation is over all triangles in $T$, $|\Delta|$
denotes the area of the triangle $\Delta $ in the surface
and the coupling constants $\kappa$ and $\lambda$
are in a suitable range.

Let us denote the partition functions for this model by
$Z_T(\kappa ,\lambda )$ and $Z(\mu ,\kappa ,\lambda )$,
defined in analogy with \rf{2} and \rf{3}, respectively.  It is
easy to see that the proof given above of the divergence of
$Z(\mu ,\lambda )$ breaks down for $Z(\mu ,\kappa ,\lambda
)$, if $\kappa $ is chosen large enough for given $\mu$ and
$\lambda $.

We do not have a proof that $Z(\mu ,\kappa ,\lambda )$ is
finite for any values of the coupling constants.  For this one
would need an exponential bound
\beq
Z_T(\kappa ,\lambda )\leq C(\kappa ,\lambda )^{|T|}\label{12}
\eeq
for some $C(\kappa ,\lambda )>0$ and $T$ of fixed topology.
A proof of \rf{12} by successive integration over vertices is
not straightforward.

Let us finally mention that provided $Z(\mu ,\kappa ,\lambda )$
is finite in some region $\Omega$ of coupling constants, one
can prove by standard methods elementary properties of the
model such as convexity of $\Omega$, non-negativity and concavity
of the mass and
the string tension as well as linear upper bounds on these
quantities (see \cite{adf,dfj,adfo,adf2,adfj2}).
In this model it appears, however, to be just as difficult to
establish the existence of a critical point at which both
the mass and the string tension scale as it is in the Gaussian
models with extrinsic curvature terms added to the the action.

\end{document}